\documentclass[%
 reprint,
superscriptaddress,
 amsmath,amssymb,
 aps,
 prl,
]{revtex4-1}

\usepackage{graphicx}
\usepackage{dcolumn, braket}
\usepackage{bm}
\usepackage{xcolor}

\begin{document}

\preprint{APS/123-QED}

\title{Unveiling and Manipulating Hidden Symmetries in Graphene Nanoribbons}

\author{Nikita V.~Tepliakov}
\email{n.tepliakov20@imperial.ac.uk}
\affiliation{Departments of Materials and Physics, Imperial College London, London SW7 2AZ, United Kingdom}
\affiliation{The Thomas Young Centre for Theory and Simulation of
Materials, Imperial College London, London SW7 2AZ, United Kingdom}

\author{Johannes Lischner}
\affiliation{Departments of Materials and Physics, Imperial College London, London SW7 2AZ, United Kingdom}
\affiliation{The Thomas Young Centre for Theory and Simulation of
Materials, Imperial College London, London SW7 2AZ, United Kingdom}
\author{Efthimios Kaxiras}
\affiliation{School of Engineering and Applied Sciences, Harvard University, Cambridge, Massachusetts 02138, United States}
\affiliation{Department of Physics, Harvard University, Cambridge, Massachusetts 02138, United States}

\author{Arash A.~Mostofi}
\affiliation{Departments of Materials and Physics, Imperial College London, London SW7 2AZ, United Kingdom}
\affiliation{The Thomas Young Centre for Theory and Simulation of
Materials, Imperial College London, London SW7 2AZ, United Kingdom}

\author{Michele Pizzochero}
\email{mpizzochero@g.harvard.edu}
\affiliation{School of Engineering and Applied Sciences, Harvard University, Cambridge, Massachusetts 02138, United States}

\date{\today}

\begin{abstract}
Armchair graphene nanoribbons are a highly promising class of semiconductors for all-carbon nanocircuitry. Here, we present a new perspective on their electronic structure from simple model Hamiltonians and \textit{ab initio} calculations. We focus on a specific set of nanoribbons of width $n=3p + 2$, where $n$ is the number of carbon atoms across the nanoribbon axis and $p$ is a positive integer. We demonstrate that the energy-gap opening in these nanoribbons originates from the breaking of a previously unidentified hidden symmetry by long-ranged hopping of $\pi$-electrons and structural distortions occurring at the edges. This hidden symmetry can be restored or manipulated through the application of in-plane lattice strain, which enables continuous energy-gap tuning, the emergence of Dirac points at the Fermi level, and topological quantum phase transitions. Our work establishes an original interpretation of the semiconducting character of armchair graphene nanoribbons and offers guidelines for rationally designing their electronic structure.
\end{abstract}

\maketitle

\smallskip
\paragraph{Introduction.} Graphene nanoribbons (GNRs) --- few-atom wide strips of $sp^2$-bonded carbon atoms --- are prime candidates for  post-silicon electronics \cite{Yazyev2013, Wang2021a} owing to the combination a sizable energy gap \cite{Han2007, Xiaolin2008, Chen2013}, intrinsic high carrier mobility \cite{Wang2021b}, long mean free path \cite{Baringhaus2014}, and peculiar field effects \cite{Son2006b, Novikov2007, Pizzochero2021a}. GNRs can be fabricated in an atom-by-atom fashion by bottom-up on-surface synthesis \cite{Zongping2020, Yano2020}, resulting in a wide spectrum of atomically precise edges \cite{Cai2010a} and more complex structures \cite{Groning2018, Sun2020a, Pizzochero2021c, Rizzo2018}. Of particular interest for next-generation logic electronics are armchair graphene nanoribbons (AGNRs) \cite{Cai2010a, Tarliz2017} and related one-dimensional heterojunctions \cite{Chen2015a, Jacobse2017, Wang2017, Pizzochero2020a}  because of their transferability onto insulating substrates \cite{BorinBarin2019}, fabrication scalability \cite{DiGiovannantonio2018}, and width-controllable energy gaps \cite{Merino2017, Chen2013, Son2006a}. Integration of AGNRs into experimental devices has led to the realization of short-channel field-effect transistors operating at room temperature that exhibit high on-to-off current ratios and on-currents at finite voltages \cite{Llinas2017, Jacobse2017, Braun2021}. 

Previously, it has been shown that the opening of an energy gap in AGNRs can be explained either by structural distortions occurring at the edges \cite{Son2006b} or by long-range hopping interactions between $\pi$-electrons \cite{gunlycke2008tight}. In this Letter, we provide a fresh understanding of the electronic structure of AGNRs by demonstrating that their energy-gap opening originates from the breaking of a previously overlooked hidden symmetry. This is in contrast to zigzag-edged nanoribbons, where the energy-gap opening is driven by electron-electron interactions \cite{Son2006a, Yazyev2010, Yazyev2011}. We show that this hidden symmetry can be manipulated by means of lattice strain and used to enforce Dirac points at the Fermi level or cause topological quantum phase transitions. Through these insights, we formulate guidelines to engineer the electronic properties of armchair graphene nanoribbons.

\begin{figure*}
    \centering
    \includegraphics[width=\textwidth]{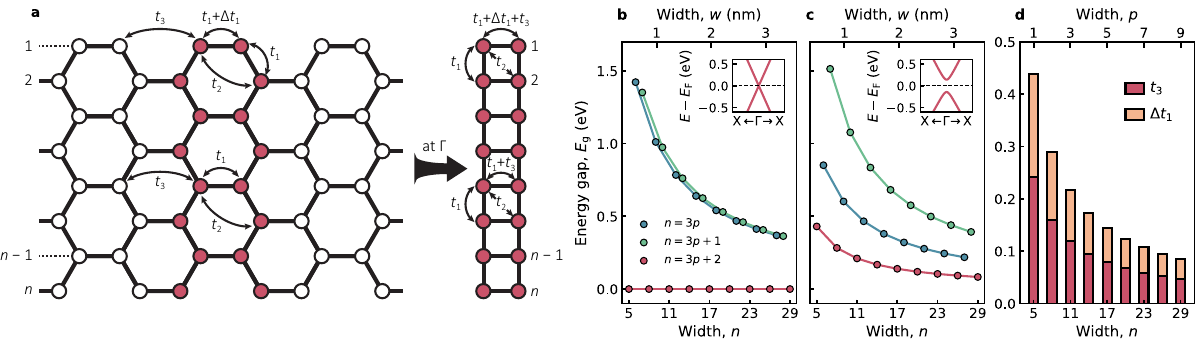}
    \caption{(a) Crystal structure of armchair graphene nanoribbons (AGNRs) with $n$ carbon atoms in the lateral direction, along with the isomorphic transformation at the center of the Brillouin zone, $\Gamma$, resulting in a finite-size auxiliary model. Also illustrated are the first, second, and third nearest-neighbor  hopping parameters $t_1$, $t_2$, and $t_3$, respectively, as well as the change of the first-nearest neighbor hopping parameter at the edges of the nanoribbon, $\Delta t_1$ due to a structural distortion. Evolution of the energy gap, $E\textsubscript{g}$, of nanoribbons of increasing width $n$ obtained using the tight-binding Hamiltonian in Eq.~(\ref{Eqn1}) when (b) $t_2$, $t_3$, and $\Delta t_1$  are ignored, and (c) $t_2$, $t_3$, and $\Delta t_1$ are included. The insets show the band structures in the vicinity of the $\Gamma$ point for the representative width $n=8$. (d) Contribution of $t_3$ and $\Delta t_1$ to $E_\mathrm{g}$ of AGNRs with width $n=3p+2$.}
    \label{Fig1}
\end{figure*}

\smallskip
\paragraph{Origin of energy gaps in AGNRs.}
To elucidate the electronic structure of AGNRs, we rely on the tight-binding Hamiltonian for the $p_z$-electrons~\cite{Pereira2009},
\begin{equation}
    \hat{\mathcal{H}} = \left( \sum_{\braket{i,j}}t_1 + \sum_{\braket{\braket{i,j}}}t_2 + \sum_{{\braket{\braket{\braket{i,j}}}}}t_3 \right) (\hat c^\dagger_i \hat c_j + \mathrm{h.c.}),
    \label{Eqn1}
\end{equation}
where $t_1 = -2.88$~eV, $t_2 = 0.22$~eV, and $t_3 = -0.25$~eV are the first, second, and third nearest-neighbor hopping interactions, respectively, as illustrated in Fig.~\ref{Fig1}(a). Importantly, the first nearest-neighbor hopping interactions at the edges of the nanoribbon are smaller than those in the inner region by $\Delta t_1 = -0.20$~eV because of the shortening of the interatomic distance between carbon atoms at the edges by $\sim$4\% as a result of a lattice relaxation~\cite{Son2006a}. We obtained the amplitudes of the hopping parameters through Wannierization of the \textit{ab initio} band structure, as described in the Supplemental Material~\cite{supplementary} (Supplemental Note~1). This \textit{ab initio}-parameterized tight-binding Hamiltonian allows us to effectively disentangle the contribution of each hopping interaction to the resulting electronic structure. 

For completeness, in Fig.~\ref{Fig1}(b,c) we start by recalling the dependence of the energy gaps of AGNRs, $E\textsubscript{g}$, on their width, which is quantified by the number of carbon atoms across the nanoribbon axis, $n$. Nanoribbons can be grouped according to their width, depending on whether $n = 3p$,  $3p+ 1$, or $3p+ 2$, where $p$ is a positive integer~\cite{Son2006a}. First, we consider only nearest-neighbor hopping interactions and ignore the shortening of the bond lengths at the edges of the nanoribbon; see Fig.~\ref{Fig1}(b). For $n = 3p+2$, AGNRs are gapless and possess a zero-energy Dirac point, while for $n = 3p$ or $n = 3p+1$ they exhibit a direct energy gap located at the center of the Brillouin zone, the magnitude of which scales inversely with the width. Next, we account for both the farther nearest-neighbor hopping interactions, $t_2$ and $t_3$, and the shortening of the bond lengths at the edges; see Fig.~\ref{Fig1}(c). A width-dependent energy gap opens in AGNRs with $n = 3p+2$ too, in agreement with earlier theoretical works \cite{gunlycke2008tight,Son2006a}. In Fig.~\ref{Fig1}(d), we assess the relative contribution of these two mechanisms to the energy gap of AGNRs, as further detailed in Supplemental Material~\cite{supplementary} (Supplemental Notes 2--3). Our analysis reveals that the effect of long-range hopping interactions and structural distortions at the edges are comparable in determining the magnitude of the energy gap, with the contribution from including $t_3$ exceeding that from accounting for $\Delta t_1$ by $\sim$20\%. Overall, this observation points to a distinct nature of the semiconducting state in AGNRs of width $n=3p+2$ compared to $n=3p$ or $n=3p+1$.

\begin{figure}[t]
    \centering
    \includegraphics[width=1\columnwidth]{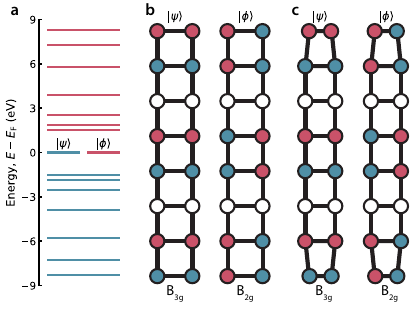}
    \caption{(a) Energy spectrum at the $\Gamma$ point of an AGNR with $n=8$, as obtained from the tight-binding Hamiltonian ignoring long-range hopping interactions and the structural distortion at the edges of the nanoribbon ($t_2 = t_3 = \Delta t_1 = 0$). (b) Wavefunctions corresponding to the pair of zero-energy degenerate levels $\ket{\psi}$ and $\ket{\phi}$ of the auxiliary model with $n=8$, as obtained through the  tight-binding Hamiltonian with $t_2 = t_3 = \Delta t_1 = 0$; red and blue colors indicate opposite signs of the wavefunction. (c) Same as panel (b), but including long-range hopping interactions, $t_2$ and $t_3$, and the structural distortion occurring at the edges of the nanoribbon, $\Delta t_1$, in the tight-binding Hamiltonian.}
    \label{Fig2}
\end{figure}

To explain the origin of the energy-gap opening in nanoribbons with $n=3p+2$, we initially consider the nearest-neighbor tight-binding Hamiltonian with $\Delta t_1 = t_2 = t_3 = 0$ along with with an isomorphic transformation from the extended nanoribbon to the auxiliary, finite-size model shown in Fig.~\ref{Fig1}(a), which features the same energy spectrum of the corresponding AGNR at the center of the Brillouin zone \cite{Son2006a}. For the illustrative case of $n=8$ ($p = 2$), the energy spectrum of the auxiliary model is given in Fig.~\ref{Fig2}(a). The pair of zero-energy degenerate levels corresponds to the Dirac point in the band structure of the equivalent AGNR.

The presence of this double degeneracy is surprising because it is not enforced by the rectangular point symmetry group D\textsubscript{2h} of the auxiliary model, which has only non-degenerate irreducible representations. These degenerate zero-energy modes, therefore, must arise from an additional, hidden symmetry. The most famous example of such a hidden symmetry is the degeneracy of the energy levels of the hydrogen atom with respect to the orbital angular momentum, associated with the Laplace-Runge-Lenz vector \cite{fock1935theorie}. To reveal the hidden symmetry associated with the auxiliary model, we solve analytically its nearest-neighbor tight-binding Hamiltonian [see Supplemental Material~\cite{supplementary} (Supplemental Note 4) for details], deriving energy levels and states
\begin{equation}
    E_{m_x m_y} = 2t_1 \left[\cos{\left(\frac{\pi m_x}{3}\right)} + \cos{\left(\frac{\pi m_y}{n + 1}\right)} \right],
\end{equation}
\begin{equation}
    \psi_{m_x m_y}(x_i,y_i) = \frac{2}{\sqrt{3(n+1)}}\sin{\left(\frac{\pi m_x x_i}{3}\right)} \sin{\left(\frac{\pi m_y y_i}{n+1}\right)},
\end{equation}
where $m_x=1,2$ and $m_y=1,2,\dots,n$ are the quantum numbers that label the states while $x_i$ and $y_i$ are the actual Cartesian coordinates of the $i$-{th} atom of the auxiliary model.

These states $\psi_{m_x m_y}$ are identical to the continuum  wavefunctions of the two-dimensional quantum box of dimensions $3\times(n+1)$, mapped onto the atoms of the auxiliary model. Notably, the rectangular quantum box possesses a hidden symmetry associated with the four-fold rotation around one of the corners of the box, $\hat C_4\psi(x,y) = \psi(y,-x)$ \cite{Lemus1998}. This symmetry enforces the seemingly accidental degeneracy of pairs of states $(m_x, m_y)$ and $(m'_x, m'_y)$ that transform into each other as $\psi_{m_x m_y} = -\hat C_4\psi_{m'_x m'_y}$. Such pairs are determined by the conditions
\begin{equation}
    m'_x = \frac{3}{n + 1} m_y,
    \quad
    m'_y = \frac{n + 1}{3} m_x.
\end{equation}
Since $m_x$ only assumes values $1$ and $2$, the ratio $(n + 1)/3$ must be an integer and we naturally obtain the condition $n = 3p + 2$, yielding
\begin{equation}
    m'_x = \frac{m_y}{p + 1} ,
    \quad
    m'_y = (p+1) m_x.
\end{equation}
For any $n$, there is only one pair of states that satisfies these conditions: $(m_x,m_y)=(1,2p+2)$ and $(m'_x,m'_y)=(2,p+1)$. Importantly, the electron-hole symmetry following from the nearest-neighbor tight-binding Hamiltonian requires that this double degeneracy occurs at the Fermi level, hence explaining the vanishing energy gap at this level of theory. Note, however, that the hidden symmetry is only relevant at the $\Gamma$ point, and that the AGNRs are gapped throughout the rest of the Brillouin zone.

\begin{figure*}[]
    \centering
    \includegraphics[width=0.95\textwidth]{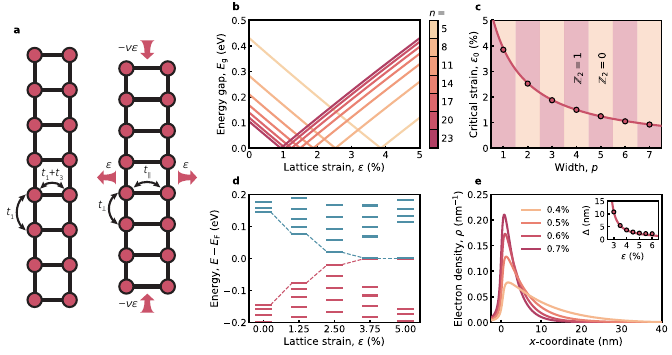}
    \caption{(a) Illustration of the lattice strain exerted on the auxiliary model, resulting in the elongation $\varepsilon$ in the periodic direction and compression $-\nu\varepsilon$ in the transverse direction, where $\nu$ is the Poisson ratio. (b) Evolution of the energy gap, $E\textsubscript{g}$, of AGNRs with $n=3p+2$ as a function of applied lattice strain, $\varepsilon$. (c) Evolution of the critical value of strain required to close the energy gap, $\varepsilon_0$, as a function of $p$; orange and red shaded areas denote trivial and topological insulating phases, respectively. (d) Energy spectrum of a finite-size AGNR with $n=8$ and length 50 nm as a function of $\varepsilon$. (e) Electron density of one of the two end states for increasing values of strain, visualized along the nanoribbon $x$-axis; the inset shows the decay length of the end states, $\Delta$, as a function of the applied strain.}
    \label{Fig3}
\end{figure*}

To further understand the effect of the hidden symmetry $\hat C_4$ on the electronic structure, we examine the wavefunctions of the two degenerate states, $\ket{\psi}$ and $\ket{\phi}$, illustrated in Fig.~\ref{Fig2}(b). As can be seen in the figure, these wavefunctions are strictly localized in regions of square symmetry that are separated by nodal lines (sites denoted by empty circles). Even though $\ket{\psi}$ and $\ket{\phi}$ belong to the irreducible representations B\textsubscript{3g} and B\textsubscript{2g} of the D\textsubscript{2h} group, the wavefunction in each square region is reminiscent of a doubly degenerate irreducible representation E\textsubscript{2g} of D\textsubscript{4h}, which is the point symmetry group of the square molecule cyclobutadiene (C$_4$H$_4$), arguably the simplest $\pi$-electron model system hosting zero-energy degenerate levels. And it is indeed the case that the wavefunctions of these degenerate states $\ket{\psi}$ and $\ket{\phi}$ transform into each other through a hidden four-fold rotational symmetry, as demonstrated in Supplemental Material~\cite{supplementary} (Supplemental Note 5); furthermore, in Supplemental Material~\cite{supplementary} (Supplemental Note 6) we show and discuss the full set of wavefunctions of the auxiliary model.

The origin of the degeneracy in the energy spectrum of the ANGRs with $n = 3p+2$, therefore, traces back to the presence of a hidden four-fold rotational symmetry $\hat C_4$, that is, a symmetry operation of the electronic states that is not inherent to the symmetry of the crystal structure. This hidden symmetry enforces the peculiar square-shaped localization pattern of electrons, which only AGNRs of the $n=3p+2$ class can accommodate according to our analysis.

The energy-gap opening at the center of the Brillouin zone arises as a consequence of long-range hopping interactions between $\pi$-electrons and structural distortions at the edges of the nanoribbon. Both mechanisms are operative in deforming the square-shaped localization pattern of the zero-energy states, as illustrated in Fig.~\ref{Fig2}(c), thus breaking the hidden symmetry. Among the long-range hopping interactions, only $t_3$ is effective in lifting the hidden symmetry through the effective shortening of the longitudinal bonds in the auxiliary model, as shown in Fig.~\ref{Fig1}(a). The second nearest-neighbor hopping $t_2$ is ineffective to this purpose as, by acting along the diagonal of the square, it preserves the four-fold rotation. While the energy gap within each square-shaped fragment is determined only by $t_3$ and $\Delta t_1$, the overall energy gap decreases with the width of the nanoribbon because the square fragments are coupled through $t_2$ and $t_3$.

\smallskip
\paragraph{Strain engineering of AGNRs.} 
Having established the origin of the semiconducting state in AGNRs of width $n=3p+2$, we next devise strategies to engineer their electronic structure. This is accomplished by exploiting the interplay between the spatial and hidden symmetries through lattice deformations. As illustrated in Fig.~\ref{Fig3}(a), tensile strain exerted along the nanoribbon axis decreases the longitudinal hopping interactions in the auxiliary model from $t_1 + t_3$ to $t_\parallel \approx (t_1 + t_3)(1 - \gamma\varepsilon)$, where $\gamma = 2.0$ is the spatial decay constant of the hopping integrals and $\varepsilon$ is the relative elongation. Due to the Poisson effect, such an elongation induces a concurrent compression of $-\nu\varepsilon$ in the perpendicular direction, where $\nu=0.2$ is the Poisson ratio \cite{kalosakas2021width}. This, in turn, leads to an increase of the lateral hopping interactions from $t_1$ to $t_\perp \approx t_1(1 + \gamma\nu\varepsilon)$.

In Fig.~\ref{Fig3}(b), we show that a full control over the energy gap of AGNRs can be achieved by modulating the ratio between $t_\parallel$ and $t_\perp$ via lattice strain. For each $n$, three regimes can be distinguished: (i) $t_\parallel > t_\perp$, where $E\textsubscript{g}$ decreases linearly, (ii) $t_\parallel = t_\perp$, occurring at a critical strain $\varepsilon_0$, where the hidden symmetry is restored, $E\textsubscript{g}$ is quenched, and Dirac points develop at the Fermi level, and (iii) $t_\parallel < t_\perp$, where the energy gap increases linearly. Similarly to the energy gap, the critical strain, $\varepsilon_0$, follows an approximate inverse dependence on the nanoribbon width, $w$, as
\begin{equation}
    \varepsilon_0(w) = \frac{\alpha}{w + \delta w},
\end{equation}
where $\alpha = 2.7$\%$\cdot$nm and $\delta w = 0.2$~nm, as extracted from the fitting given in Fig.~\ref{Fig3}(c). Earlier theoretical works predicted an analogous strain-dependence of the energy gaps of AGNRs \cite{Li2010, Sun2008}, although the origin of these observations remained poorly understood. Our results provide an explanation to these otherwise elusive effects by connecting them to a hidden symmetry in the electronic structure of the nanoribbons.


Finally, to assess how the observed evolution of the energy gap as a function of strain affects the band topology of AGNRs, we determine the topological invariant, $\mathbb Z_2$, as \cite{Cao2017}
\begin{equation}
    (-1)^{\mathbb Z_2} = \prod_i \delta_i,
\end{equation}
where $\delta_i = \pm 1$ is the parity of the $i$th band with respect to reflection across the nanoribbon width ($x\to-x$, where $x$ is the periodic direction), with the product taken over all occupied bands. According to this definition, an even (odd) number of antisymmetric valence states yields  $\mathbb Z_2 = 0$ ($\mathbb Z_2 = 1$), denoting a trivial (non-trivial) band topology. For unstrained AGNRs, $(-1)^{\mathbb Z_2} = (-1)^n$, so that the topology is strictly defined by the width, i.e., nanoribbons with an even (odd) $n$ are topologically trivial (non-trivial) \cite{Cao2017, Lopez2021}.

Fig.~\ref{Fig3}(c) shows the dependence of $\mathbb Z_2$ on the strength of the lattice deformation. The energy-gap closing and subsequent reopening driven by strain is accompanied by a change in ${\mathbb Z_2}$ between 0 and 1. This signals that topologically trivial nanoribbons become non-trivial, and vice versa. The observed topological quantum phase transition stems from the fact that the two states that form the band edges [cf.~Fig.~\ref{Fig2}(c)] have opposite parities with respect to reflection across the width of the nanoribbon. When the strain 
is strong enough to exchange the energy ordering of these two bands, the total parity of the occupied bands flips and so does the topological invariant.

We further analyze the emergence of the topological phase transition by considering a finite-length nanoribbon.  In Fig.~\ref{Fig3}(d), we present the effect of lattice strain on the energy spectrum of a representative topologically trivial finite-size ANGR of width $n=8$ and length of 50~nm. The application of strain continuously reduces the energy gap, which eventually vanishes at the critical value $\varepsilon_0 = 2.5$ \%, giving rise to a pair of zero-energy degenerate states. These are topologically protected end states which, as shown in Fig.~\ref{Fig3}(e), selectively reside at the nanoribbon ends and are the hallmark of one-dimensional topological insulators \cite{Rizzo2018, Rizzo2021, Groning2018}. Their electron density decays exponentially into the nanoribbon, $\rho(x) \propto e^{-x / \Delta}$, where $\Delta$ is the decay length. By fitting, we obtain that the decay length of these end states depends on strain according to the relation
\begin{equation}
    \Delta = \frac{\Delta_0}{|\varepsilon - \varepsilon_0|},
\end{equation}
where $\Delta_0 = 5.4\%\cdot$~nm. Such end states can be experimentally probed through scanning tunneling spectroscopy~\cite{Lawrence2020}.

\smallskip
\paragraph{Summary and conclusions.} 
In summary, we have demonstrated that the electronic structure of armchair graphene nanoribbons of width $n=3p+2$ is dominated by a previously unidentified hidden symmetry. The energy-gap opening originates from the hidden-symmetry breaking induced by long-range hopping interactions between $\pi$-electrons and structural distortions occurring at the edges of the nanoribbon. The identification of this hidden symmetry enables the engineering of the properties of graphene nanoribbons through lattice deformations, which can give rise to Dirac points at the Fermi level, tunable energy gaps, and topological quantum phase transitions. To conclude, our findings present a novel interpretation of the semiconducting character of armchair graphene nanoribbons and provide means for the precise control of their functionalities.

\smallskip
\paragraph{Acknowledgments.} 
N.V.T.~acknowledges the President's PhD Scholarship of Imperial College London. M.P.\  is financially supported by the Swiss National Science Foundation (SNSF) through the Early Postdoc.Mobility program (Grant No.\ P2ELP2-191706) and the NSF DMREF (Grant No.\ 1922172).

%


\end{document}